\documentclass[12pt]{article}

\usepackage{latexsym}
\usepackage{epsfig}

\usepackage{graphicx}
\usepackage{epstopdf}

\usepackage{epsfig,amssymb,amsmath,amscd}

\newcommand{\bq}{\begin{equation}}
\newcommand{\eq}{\end{equation}}
\newcommand{\bqa}{\begin{eqnarray}}
\newcommand{\eqa}{\end{eqnarray}}
\newcommand{\ben}{\begin{enumerate}}
\newcommand{\een}{\end{enumerate}}
\newcommand{\bc}{\begin{center}}
\newcommand{\ec}{\end{center}}
\newcommand{\bqb}{\begin{eqnarray*}}
\newcommand{\eqb}{\end{eqnarray*}}

\def\pr#1#2#3{Phys. Rev. ${\bf{#1}}$, #2 (#3)}

\def\pl#1#2#3{Phys. Lett. ${\bf{#1}}$, #2 (#3)}

\def\np#1#2#3{Nucl. Phys. ${\bf{#1}}$, #2 (#3)}

\def\jhep#1#2#3{JHEP ${\bf{#1}}$, #2 (#3)}

\def\lnc#1#2#3{Lett.Nuov.Cim. ${\bf{#1}}$, #2 (#3)}
\def\jmp#1#2#3{J. Mod. Phys. ${\bf{#1}}$, #2 (#3)}

\begin{document}
\pagenumbering{arabic}
\thispagestyle{empty}
\def\thefootnote{\fnsymbol{footnote}}
\setcounter{footnote}{1}

\begin{flushright}
Jan.10, 2017\\
arXiv: 1701.03382\\
 \end{flushright}

\vspace{2cm}

\begin{center}
{\Large {\bf  $t_{L,R}$ inclusive distributions as tests of top compositeness}}\\
 \vspace{1cm}
{\large F.M. Renard}\\
\vspace{0.2cm}

Laboratoire Univers et Particules de Montpellier,
UMR 5299\\
Universit\'{e} Montpellier II, Place Eug\`{e}ne Bataillon CC072\\
 F-34095 Montpellier Cedex 5, France.\\
\end{center}

\vspace*{1.cm}
\begin{center}
{\bf Abstract}
\end{center}

We consider the inclusive top quark distributions in $e^+e^-\to t_{L,R}+\rm{anything}$.
We analyze the modifications of the basic SM contributions which would result from
$t_{R}$ compositeness for example from the presence of form factors in the $t_{R}$ couplings
and of an effective top mass $m_t(s)$.
We also look at parton like contributions representing additional new particle
production related to $t_{R}$ constituents. One observes a large sensitivity of the
$t_{R}$ inclusive distribution to these various effects with specific shapes differing from
those of the $t_{L}$ distribution and we show that these effects may even be also observable
in the unpolarized top case.

\vspace{0.5cm}
PACS numbers: 12.15.-y, 12.60.-i, 14.65.Ha, 14.80.-j; composite models.

\def\thefootnote{\arabic{footnote}}
\setcounter{footnote}{0}
\clearpage

\section{INTRODUCTION}

Compositeness is an appealing possibility for explaining
the peculiar features, especially the broad spectrum, of the fermionic
masses, see for example \cite{comp,Hcomp2,Hcomp3,partialcomp,Hcomp4}.
The very heavy top quark  $t_R$ should be especially concerned.\\
In a previous paper we have shown that many simple signals of  
$t_R$ compositeness could be found in various processes  \cite{paptR}.\\
In the present short paper we propose a global test of $t_R$ compositeness
without having to make detailed and complete process identifications.
We suggest to look at the inclusive distribution $e^+e^-\to t + \rm{anything}$
and possibly at the polarized cases
$e^+e^-\to t_L + \rm{anything}$ and $e^+e^-\to t_R + \rm{anything}$.
The shape of the top momentum distribution is due to the associated multibody production.
The power of such an inclusive distribution is that it may reveal the presence of new effects or of new particle
production, in particular invisible states, without having to observe them
explicitly.\\
In SM the leading terms are
$e^+e^-\to t\bar b W^-, t\bar t\gamma, t\bar tZ, t\bar t H$.
We illustrate their corresponding specific shapes of the top momentum distribution.
We then study the modifications of these distributions due to $t_R$ compositeness
which could appear  through a form factor in the right-handed couplings,  an effective scale dependent top mass and new multibody production related to $t_R$ constituents which may be globally described by a parton like model.
These modifications lead to small effects on the $t_L$ distribution, but to large effects on the $t_R$ one,
with sizes and shapes specific of their origin. These effects may even be observable in the unpolarized $e^+e^-\to t + \rm{anything}$ case.

Contents: Section 2 is devoted to the SM top inclusive contributions, Section 3 to the parametrization
of  $t_R$ compositeness effects (form factors, effective mass and additional partonic contribution),
Section 4 to the illustration of these effects in $t_L$, $t_R$ and unpolarized inclusive
distributions. Results are summarized in Section 5.\\

\section{BASIC SM CONTRIBUTIONS TO TOP INCLUSIVE DISTRIBUTIONS IN
$e^+e^-$ COLLISION}

We consider the inclusive distribution

\bq
{d\sigma\over dxdcos\theta}
\eq
\noindent
where $x={2p\over\sqrt{s}}$ is the reduced top momentum,
for fixed $\theta$ angle with respect to the $e^-$ direction; $s=q^2=(p_{e^+}+p_{e^-})^2$.\\
We will separately discuss the shapes of the $x$ ditributions for $e^+e^-\to t_L+\rm{anything}$,
for $e^+e^-\to t_R+\rm{anything}$ and for the unpolarized case $e^+e^-\to t+\rm{anything}$.
We ignore the 2-body $e^+e^-\to t\bar t$ contribution located at the end of the distribution, at
$x=\sqrt{1-{4m^2_t\over s}}$.\\

The polarized $t_{L,R}$ distributions will refer to the helicity states $\lambda_t=\mp1$.
The difference with the chirality states corresponding to $P_{L,R}={1\mp\gamma^5\over2}$
appearing in the expressions of the couplings is due to
mass terms suppressed at high energy but not negligible at low energy.\\

The basic tree level SM processes consist of four 3-body processes $e^+e^-\to t\bar b W^-$,
$e^+e^-\to t\bar t \gamma$, $e^+e^-\to t\bar t Z$
and $e^+e^-\to t\bar t H$. In this first exploration we do not
include higher multibody production or loop contributions. We just want to characterize the dominant shapes of the inclusive distributions.\\
We successively compute, see ref.\cite{paptR}, separately for $t_L$ and for $t_R$, the contributions of each of these processes.\\

a) $e^+e^-\to t\bar b W^-$ with four different diagrams :
$e^+e^-\to \gamma,Z \to t\bar t $ followed by $\bar t \to \bar b W^-$;
$e^+e^-\to \gamma,Z \to b\bar b $ followed by $b \to t W^-$;
$e^-\to W^- +\nu$ followed by $e^+\nu \to\to  t\bar b $; and  
$e^+e^-\to \gamma,Z \to W^-W^+$ followed by $W^+\to t\bar b$.\\

b) $e^+e^-\to \gamma,Z \to t\bar t \gamma$ with four diagrams corresponding to $\gamma$ emission by
initial $e^{\pm}$ lines or  
by final $t,\bar t$ lines in $e^+e^-\to t\bar t$ (cuts on low energy and angle will be imposed on
the photon).\\

c) $e^+e^- \to \gamma,Z \to t\bar t Z$ with the similar above four diagrams and one more with
$e^+e^- \to Z \to  ZH $ followed by $H\to t\bar t$.\\

d) $e^+e^-\to t\bar t H$ with three diagrams:
$e^+e^-\to \gamma,Z \to t\bar t $ followed by $t \to tH$ or by $\bar t \to \bar tH$;
$e^+e^- \to Z \to  ZH $ followed by $Z\to t\bar t$.\\

These SM contributions to the $t_L$, $t_R$ and unpolarized $t$
inclusive distributions are illustrated, for $\sqrt{s}=4$ TeV, in Figure 1a,b,c, respectively.\\

For $t_L$, in Figure 1a, one sees that the  $W$, $\gamma$ ,$Z$, $H$ production processes
(the above a-d cases) contribute in a respective decreasing order, their size being controlled by their basic
couplings.\\

For $t_R$, in Figure 1b, the ordering is similar except for the fact that the W contribution (with pure left couplings, its $t_R$  contributions only coming from mass terms) is now weaker than
the $\gamma$ one and comparable to the $Z$ one.\\

Adding both $t_L$ and $t_R$ one gets the unpolarized distribution with a similar ordering
as in the left case but with slightly different respective sizes as shown in Figure 1c.\\

The shape of these distributions (their increase for $x \to 1$) corresponds to
the decrease of $(q-p)^2$ apparing in the propagator of the virtual particle
associated to the top quark of momentum $p$, the total
$e^+e^-$ momentum being denoted by $q$.\\

\section{$t_R$ COMPOSITENESS EFFECTS}

In the same spirit as in \cite{paptR} we do not consider the possibility of anomalous couplings
which would generate trivial differences with the SM case, but the presence of
a form factor generated by $t_R$  compositeness which would
affect the $s$ dependency of the $t_R$ couplings to gauge and Higgs bosons at high energy (above
some new physics scale) but which would preserve the SM properties at low energy.\\
In the case of gauge bosons we use the couplings
\bq
\bar u(t)\gamma^{\mu}(g^L_VP_L+g^R_VP_R)v(\bar t)
~~\to ~~\bar u(t)\gamma^{\mu}(g^L_VP_L+g^R_VP_RF_R(s))v(\bar t)
\eq  
\noindent
where $F_R(s)$ is the $t_R$ compositeness form factor.
The $t_L$ is kept elementary with its SM point-like coupling.\\
In the numerical illustration we will use the "test-form factor"
\bq
F_R(s)={4m^2_t+M^2\over s+M^2}
\eq
which is equal to 1 at threshold and tends to 0 at high energy, $M$ being
a new physics scale taken as 0.5 TeV in the illustrations. This is just an
arbitrary choice in order to test the sensitivity of the inclusive distributions
to such modifications. Compositeness would certainly generate more involved
s dependencies.\\

This affects the diagrams where a (virtual or real) photon or Z is connected
through right-handed couplings to a top line. The $W$ (pure left-handed) couplings are not affected.
Note that final $t_L$ states can nevertheless be slightly affected by right-handed couplings contributing through mass terms.\\

 In the case of the Higgs boson with its scalar chirality violating coupling constant
\bq
g_{Htt}=-~{em_t\over2s_Wm_W}
\eq
top compositeness suggests (see \cite{paptR}) to replace the fixed value of the $m_t$
top mass by an effective mass $m_t(s)=m_tF_R(s)$ (a kind of scale dependent mass in a way similar
to what appears in QCD, but much more violent, being due to the constituent structure).\\
This will generate modifications not only to $H$ couplings but to all top mass terms,
in particular the important ones which remain after the cancellation
of the badely behaved parts of the longitudinal gauge boson amplitudes (which can be
localized using the equivalence \cite{equivalence1,equivalence2}
of $Z_L, W^{\pm}_L$ with the goldstone bosons $G^0,G^{\pm}$ which are indeed coupled
proportionally to the top mass).\\

\underline{New particle production and parton picture}\\

At high energy the inclusive process $e^+e^-\to t_R+anything$ should involve among
$anything$ the set
of new states related to the constituent contents of $t_R$. It may start (by analogy
with the hadronic $N^*$ resonances) with a set of individual $e^+e^-\to t_R+\bar t^*$
contributions; but this will only produce peaks localized at
$x\simeq 1-{m^{*2}\over s}$.\\
One can then consider the production of new  $X$ particles emitted
by top lines, $e^+e^-\to t+\bar t+X$, as well as the production of new particles (like $e^+e^-\to Z'S$, where $S$ denotes a new neutral scalar) one of them decaying into $t\bar t$,
see next section.
Again by analogy with the proton case the sum of all these
new contributions may be globally described by a parton-like picture.
The first step consists in the production of new constituents (i), for example through couplings to
photon or Z, with the cross section  ${d\sigma_i\over dcos\theta}$.
The second step is the fragmentation into a $t_R$ and the whole set of associated
new states. This leads to the inclusive partonic type cross section

\bq
{d\sigma\over dxdcos\theta}=\Sigma_i{d\sigma_i\over dcos\theta} D_i(x)
\eq

We make an arbitrary illustration choosing top-like fermion constituents for
${d\sigma_i\over dcos\theta}$ and
a normalized fragmentation function
\bq
\int^1_{1-{M^2\over s}} xD(x)dx=1
\eq
with
\bq
D(x)={6\over(1-{M^2\over s})^3}(1-x-{M^2\over s})
\eq
which favours the low x domain ($x<1-{M^2\over s}$
corresponding to a set of new states with a mass larger than $M$).
From such a contribution one therefore expects a strong modification
of the shape of the inclusive distribution.\\

\section{ILLUSTRATIONS OF POLARIZED AND UNPOLARIZED TOP DISTRIBUTIONS}

We now discuss the observable consequences of these three types of $t_R$ compositeness effects:
modified $t_R$ couplings, effective top mass and parton-like contribution.\\
We first examine the sensitivity of each of the four SM processes ((a)-(d) considered in Section 2) to modified $t_R$ couplings and to the effective top mass.\\
Concerning the $t_L$ inclusive distribution, the main ($W,\gamma,Z$) contributions
are almost not affected; only the $H$ one can be affected by these modifications but
as it contributes very little (see Figure 1) the total is almost not not modified.\\
On the opposite the $t_R$ inclusive distribution, as one can see in Figure 2a,b,c,
is strongly affected by the form factor modifying the $t_R$ couplings, by the
effective top mass and when both of them are applied.\\
In Figure 3a,b,c we show the resulting effects on the unpolarized $t$
distribution which is quantitatively modified in an observable way with respect to the SM prediction
of Figure 1c.\\
We then look at the effect of additional $X'$ production.\\
In Figure 4a we show the kinematical shapes of 3 individual contributions due to new $X_{1,2,3}$ particles with high masses ($m_{X}=1,2,2.5$ TeV respectively) emitted by the top line (producing the $t+\bar t +X$ final state) and we compare them
to the SM $t_R$ shape.\\
In Figure 4b we draw the sum of the above three contributions to a new $Z'S$ production
contributing through the processes $e^+e^-\to V \to Z'+ S\to Z't\bar t$
and $e^+e^-\to V \to S+Z'\to St\bar t$, where $S$ is a massive scalar and $Z'$ a higher $Z$ type vector boson, and we compare it to the parton-like term.
These various shapes should be considered as pure "kinematical shapes". They only correspond to phase space distributions with arbitrary
normalizations (chosen in order to give comparable sizes) and not to precise models
which may contain further effects due to precise intermediate states or resonances.
One sees that increasing $X$ masses leads to distributions located in decreasing
$x$ domains. As expected, these shapes differ from the SM one which increases
with x. The sum of these three individual contributions is indeed, in average,
rather similar to the parton-like distribution. But a single $Z'S$ term (illustrated with $m_S=m_{Z'}=0.5$ TeV) would give a somewhat different $x$ distribution.\\
For the unpolarized $t$ case, we finally compare, in Figure 4c, these three types of
individual contributions and in Figure 4d, the corresponding shapes when the SM is added
($SM+X_1+X_2+X_3$, $SM+Z'S$ and $SM$+parton).
One sees that the essential differences are localized around low $x$ values.\\

We finally look at the resulting effects of the above different aspects of $t_R$ compositeness on the complete inclusive distributions.\\
Figure 5a compares the total $t_R$ distributions corresponding to the various compositeness effects
showing large differences in sizes and shapes in particular the completely different
parton contribution (globally representing the production of new states) .\\
Figure 5b makes the same comparisons in the unpolarized case.
The first four cases lead to moderate modifications but the additional parton
contribution leads to more important new contributions at low $x$.\\

\section{CONCLUSIONS}

In this paper we have looked at $t_R$ compositeness effects on the inclusive
distributions $e^+e^-\to t_{L,R}+\rm{anything}$. We have shown that the
shape of the $e^+e^-\to t_{R}+\rm{anything}$ distribution
is very sensitive to form factor effects in the right-handed couplings,
to the occurence of a scale dependent effective top mass and to the presence  
of additional contributions due to new particle production in the "anything" that
we can summarize by a parton-like fragmentation function.\\
We have illustrated how the basic  SM contributions to this $t_{R}$ distribution due to
$e^+e^-\to t\bar b W^-, t\bar t\gamma,
t\bar tZ, t\bar t H$ production are individually affected by the above modifications
and how the new particle production and the parton-like contribution could modify
the shape of the distribution especially at low $x$.
On the opposite the $e^+e^-\to t_{L}+\rm{anything}$ distribution is almost not
modified  by these new effects.\\
As a result the unpolarized $e^+e^-\to t+\rm{anything}$ distribution is only
moderately affected by the right-handed form factor and the effective top mass
but may be nevertheless notably affected at low $x$ by the presence of new massive particles, the sum of them being possibly described by a parton-like distribution typical of a compositeness structure.\\

Our illustrations correspond to arbitrary examples of $t_R$ compositeness effects
which would modify the shapes of the inclusive distributions expected from the
SM processes. Their purpose was to show which types of experimental observations
could detect such effects.\\

This analysis of  $e^+e^-$  processes was done assuming that the new physics scale lies in the energy range of the considered collider, for ILC see \cite{Kolk} and its refs.(5,6,7), and for top physics see \cite{Vos}.
If this is not the case, other processes in hadronic collisions could be considered,
see for ex. \cite{Richard}, but more involved phenomenological and experimental analyses would be required.\\

After completion of our work, we were informed that a study of the
 effect of $t_R$ compositeness at Tevatron and LHC through an effective
four quark operator had been done in \cite{Tait}.\\

\newpage

\begin{figure}[p]
\[\hspace*{-1.8cm}
\epsfig{file=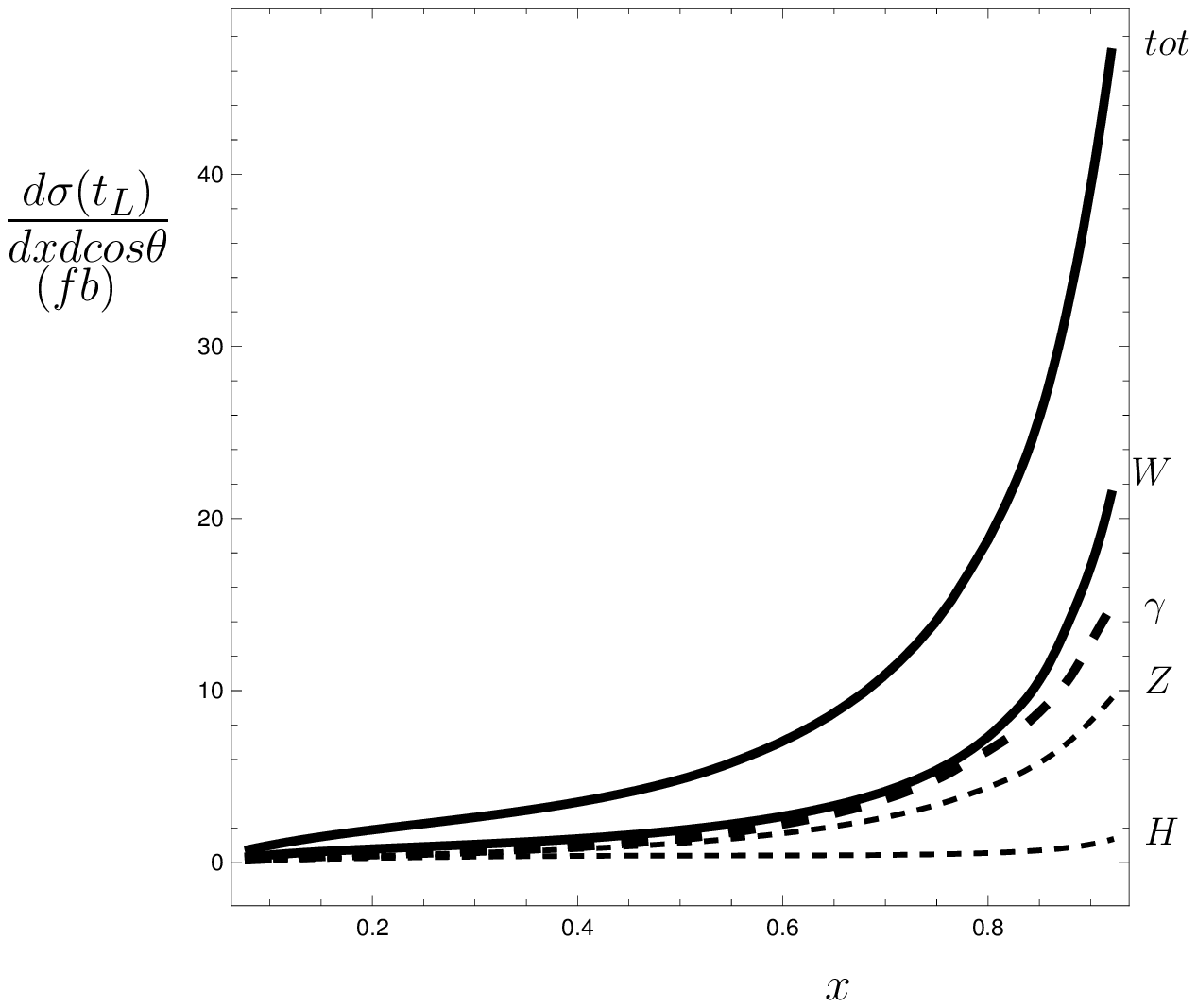, height=7cm}
\hspace*{1cm}\epsfig{file=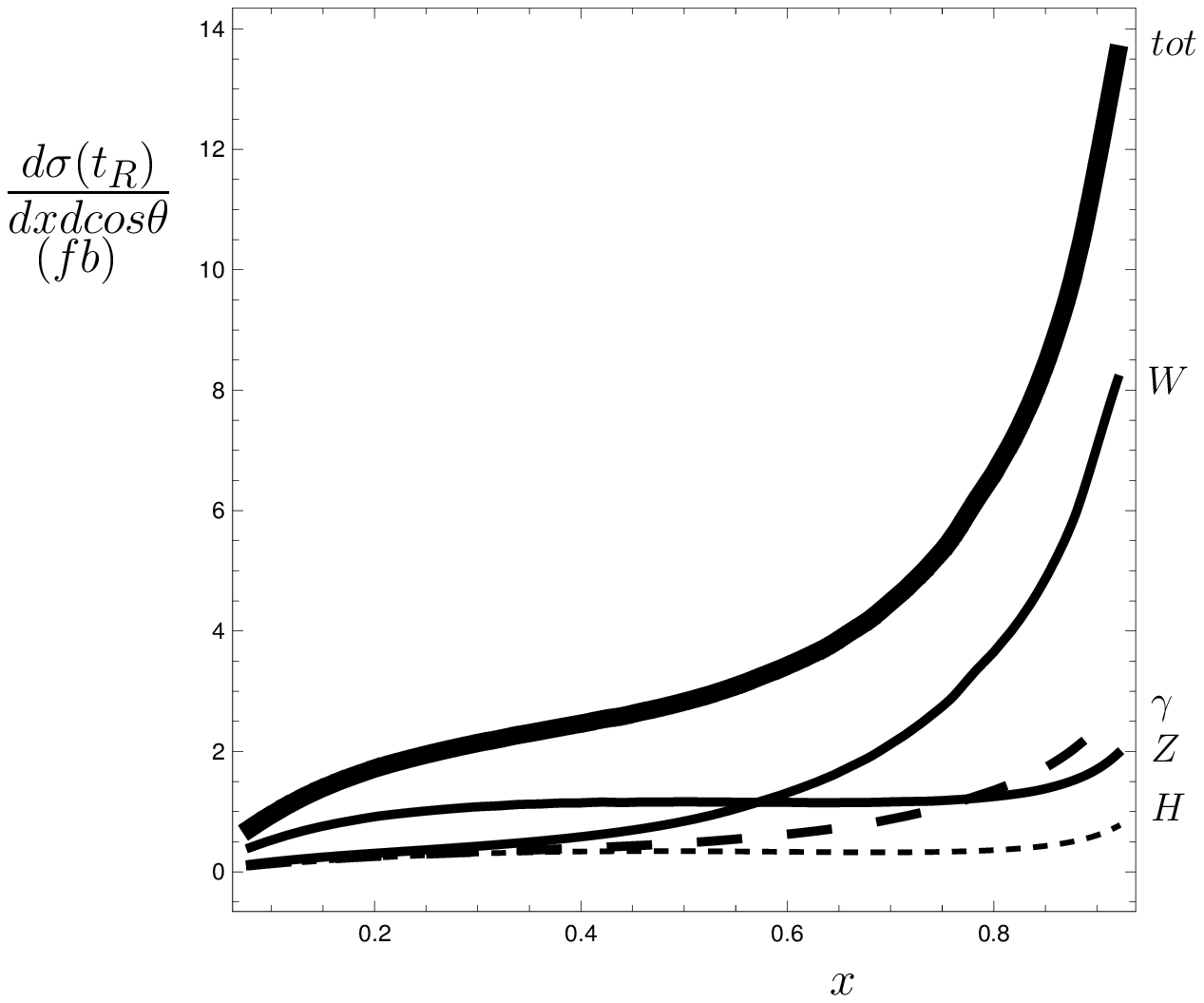, height=7cm}
\]\\
\[
\epsfig{file=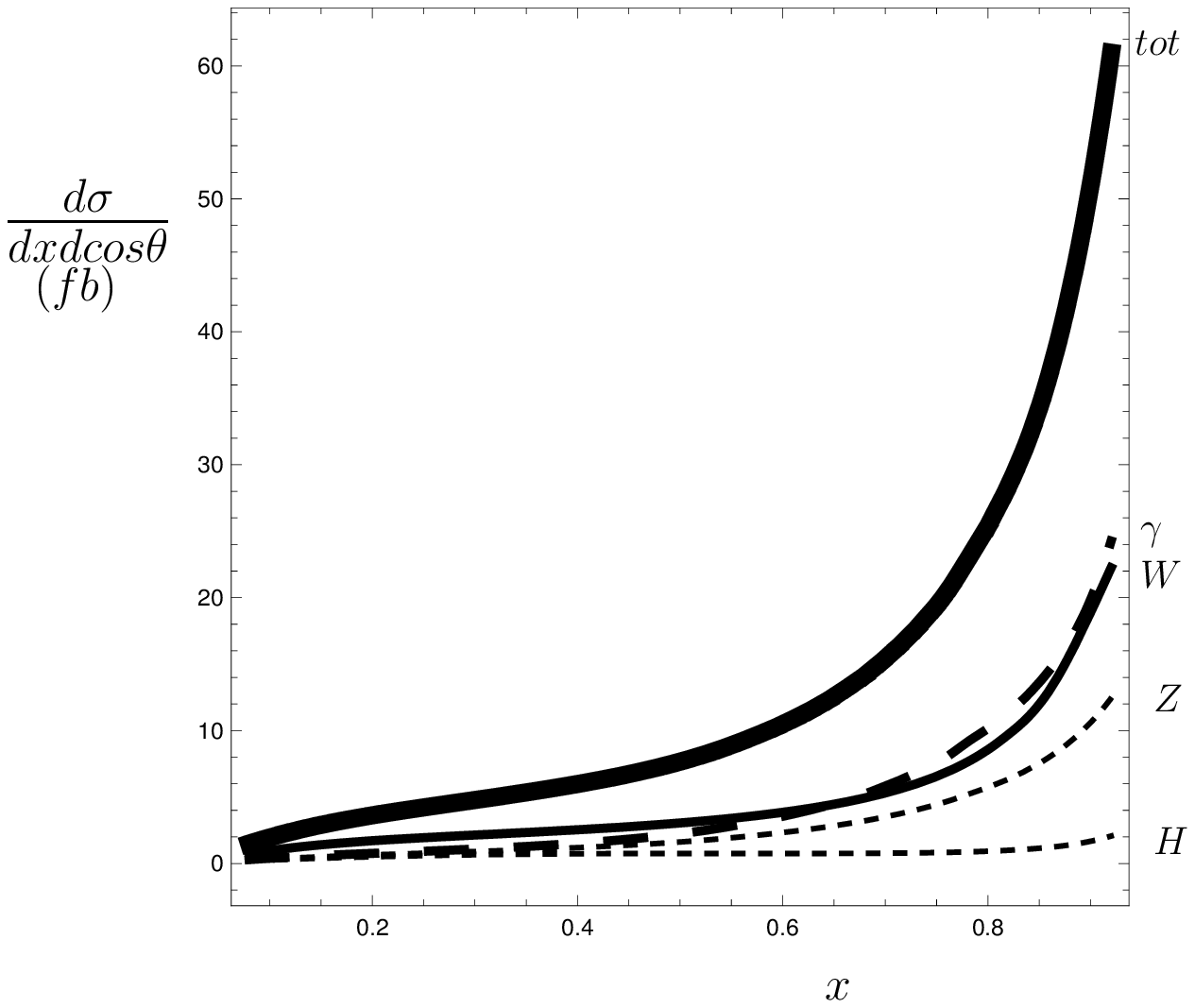, height=10.cm}
\]\\
\vspace{-1cm}
\caption[1] {Inclusive $e^+e^-\to t+\rm{anything}$ distributions due to W, $\gamma$, Z, H emission
in SM and their total; upper panel (a) for $t_L$  and (b) for $t_R$; lower panel (c) for unpolarized $t$.}
\end{figure}

\clearpage

\begin{figure}[p]
\[\hspace*{-1.8cm}
\epsfig{file=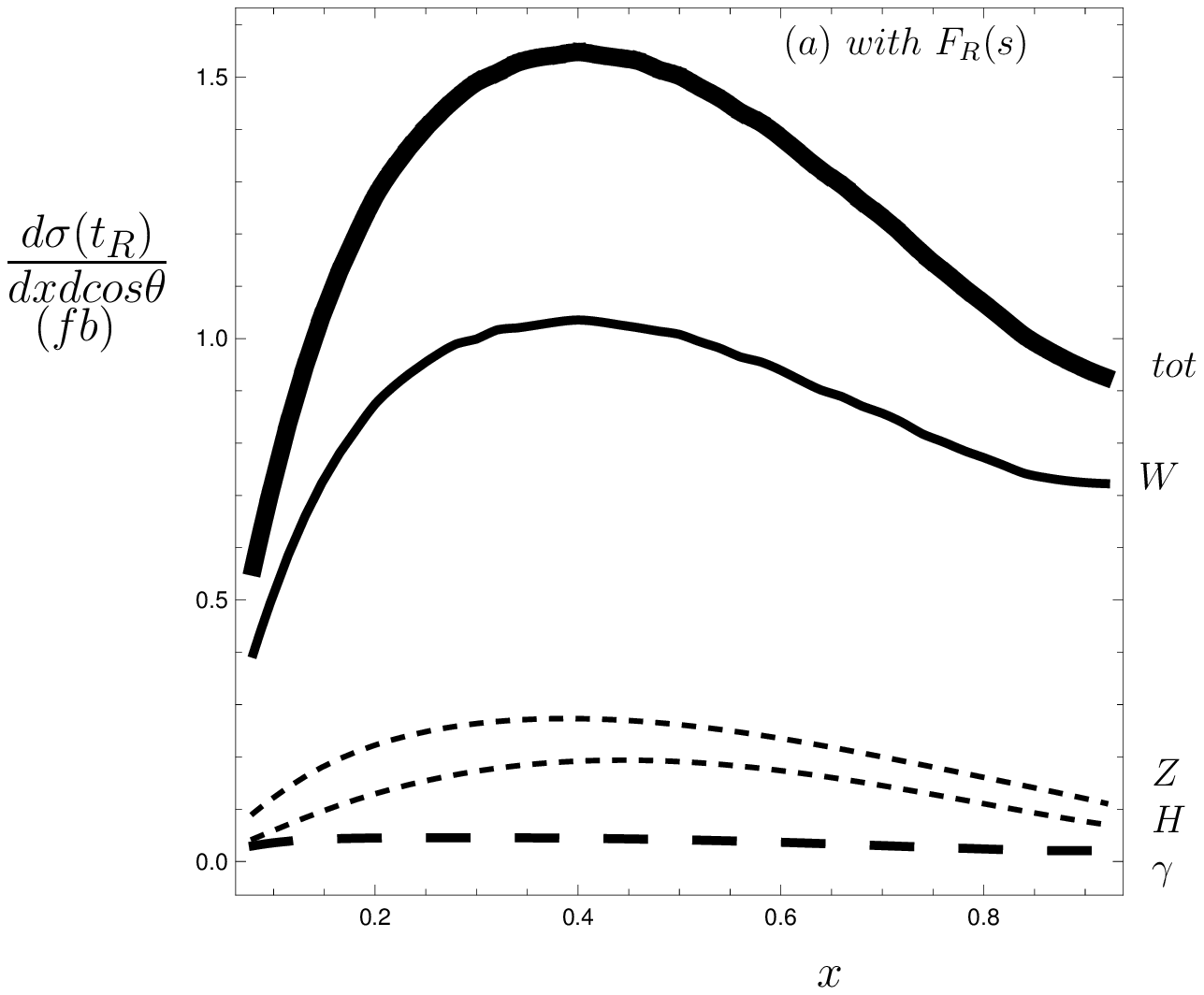, height=7cm}
\hspace*{1cm}\epsfig{file=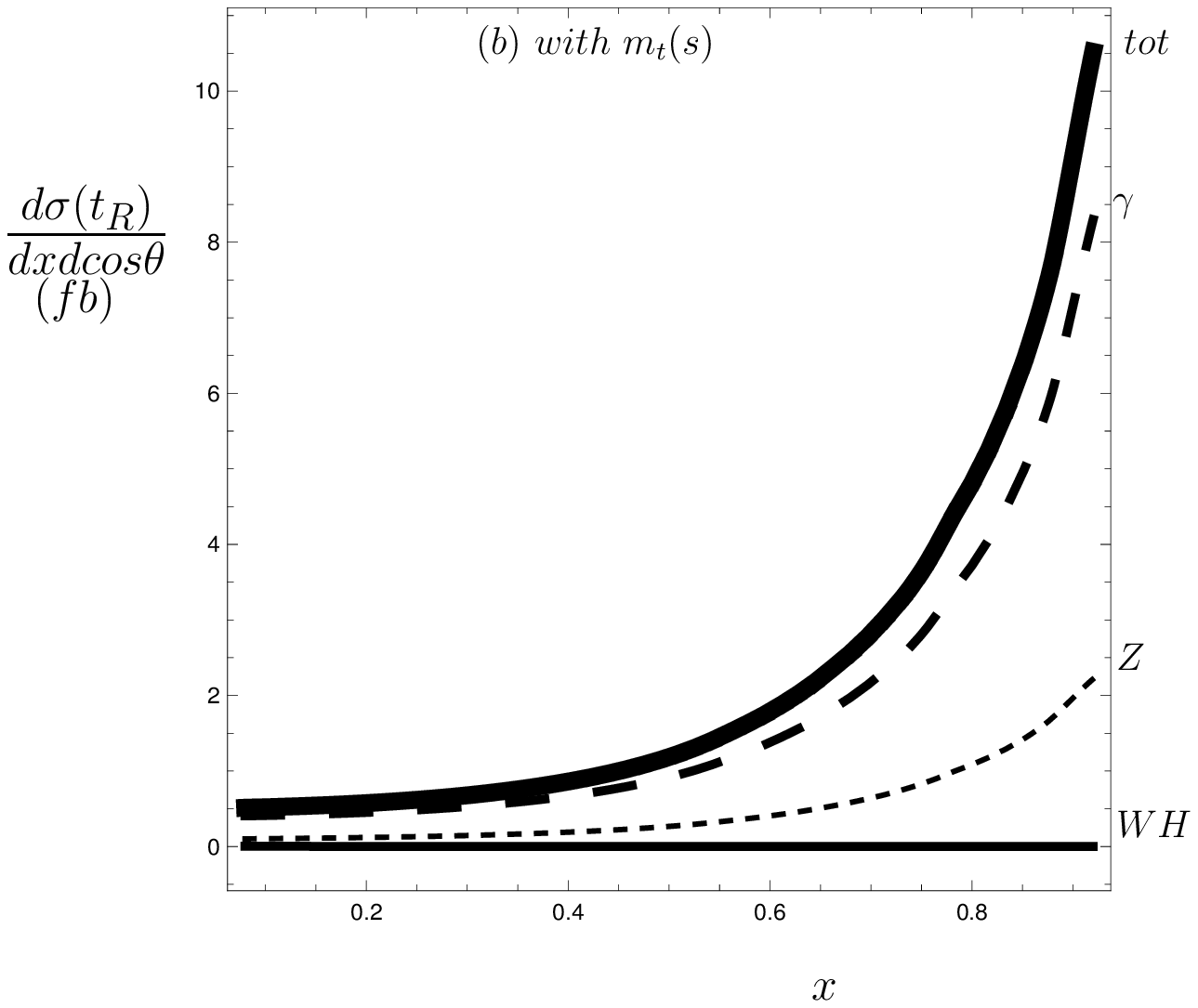, height=7cm}
\]\\
\[
\epsfig{file=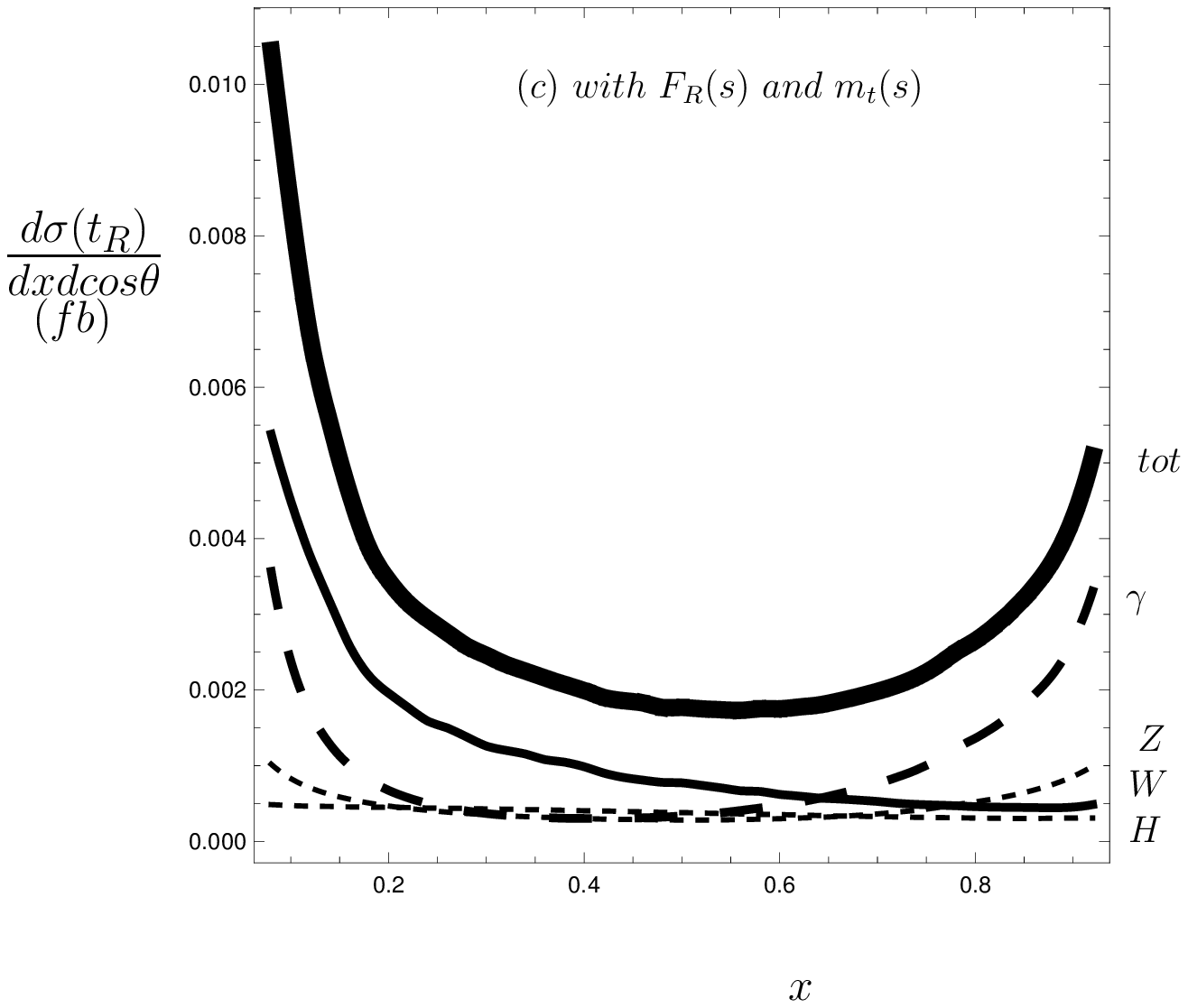, height=10.cm}
\]\\
\vspace{-1cm}
\caption[1] {Same as in Fig. 1 for $t_R$ inclusive cross section
with effects of $F_R(s)$ (a), of $m_t(s)$ (b) in the upper panel and of both (c) in the lower panel.}
\end{figure}

\clearpage

\begin{figure}[p]
\[\hspace*{-1.8cm}
\epsfig{file=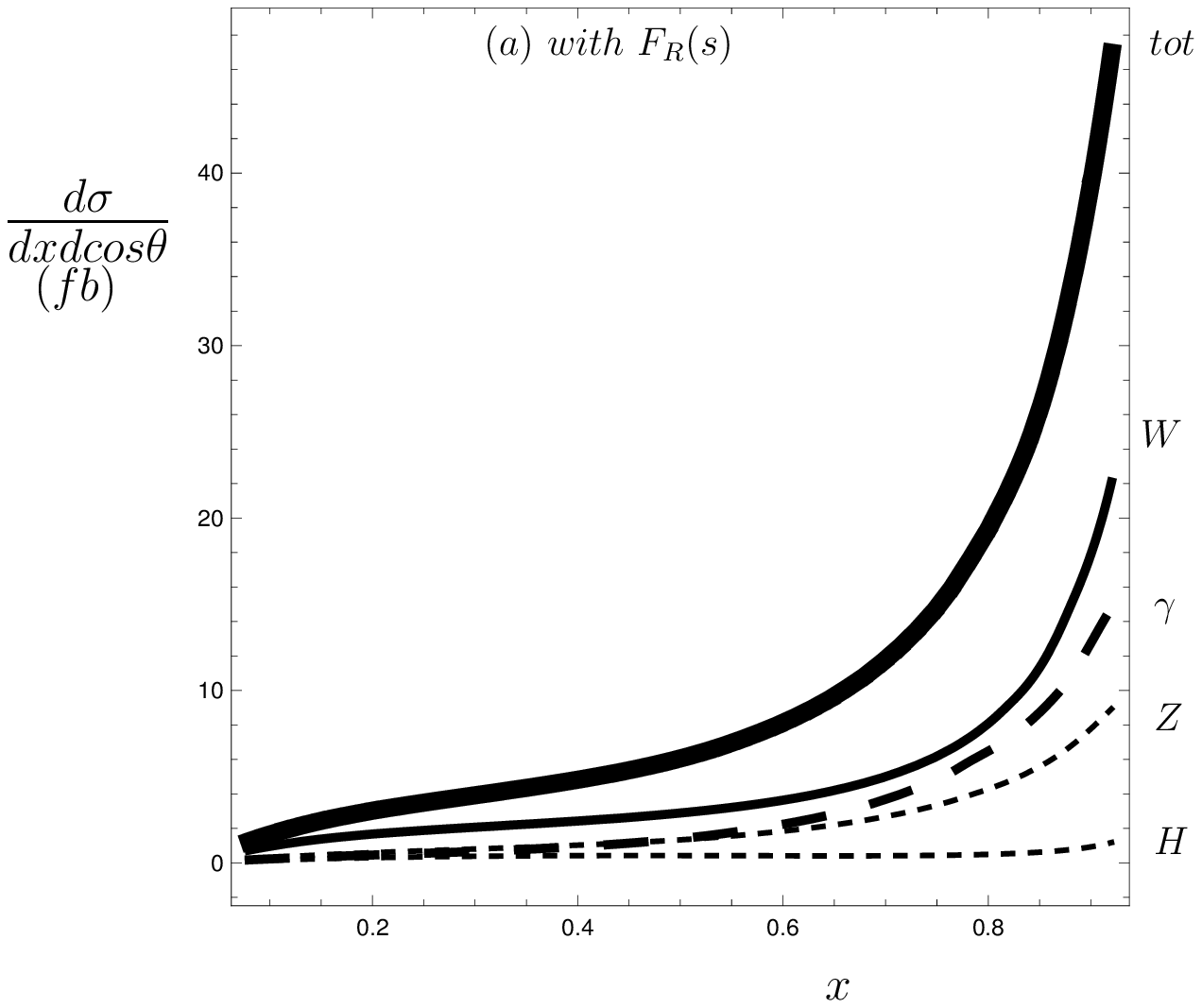, height=7cm}
\hspace*{1cm}\epsfig{file=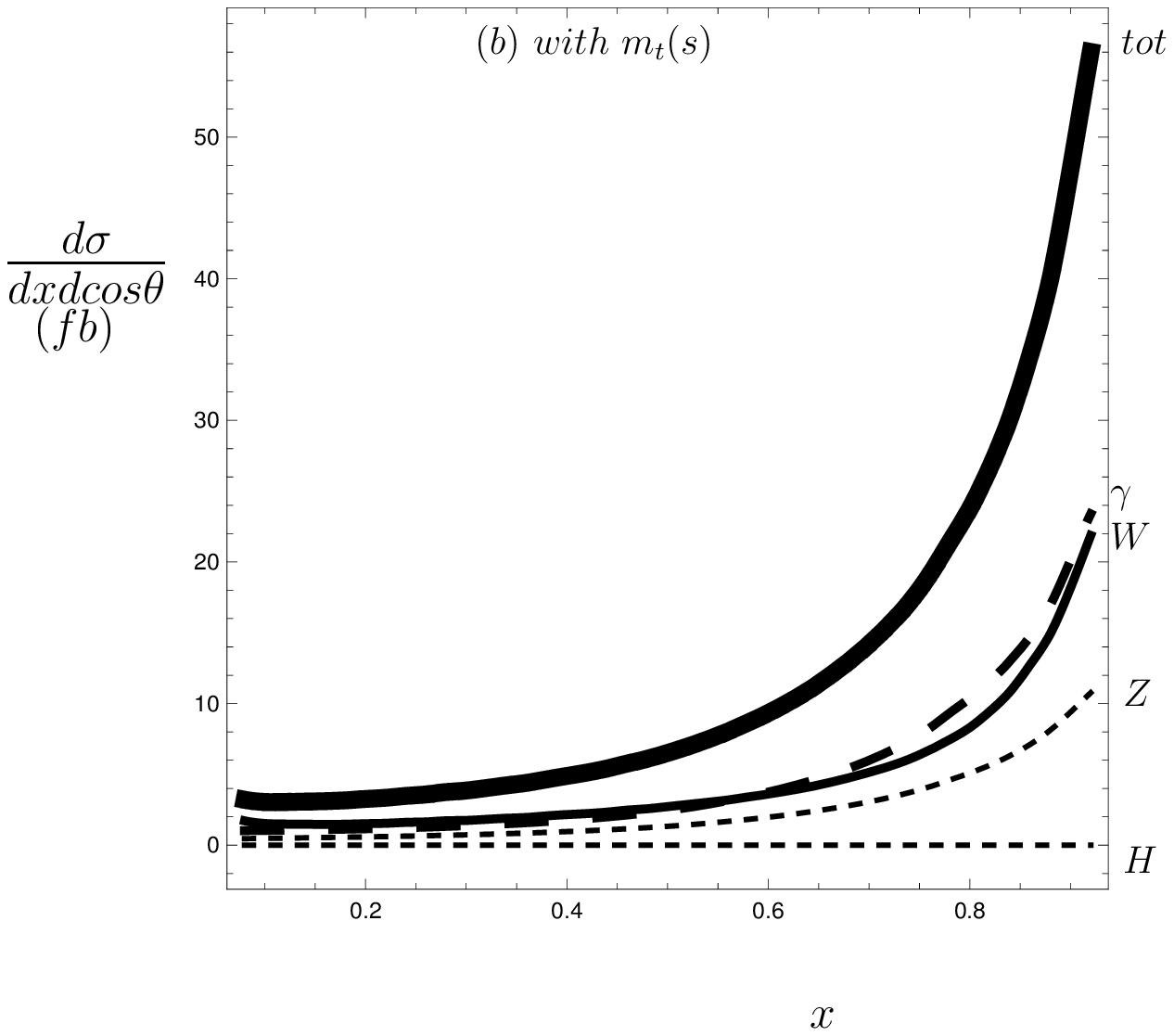, height=7cm}
\]\\
\[
\epsfig{file=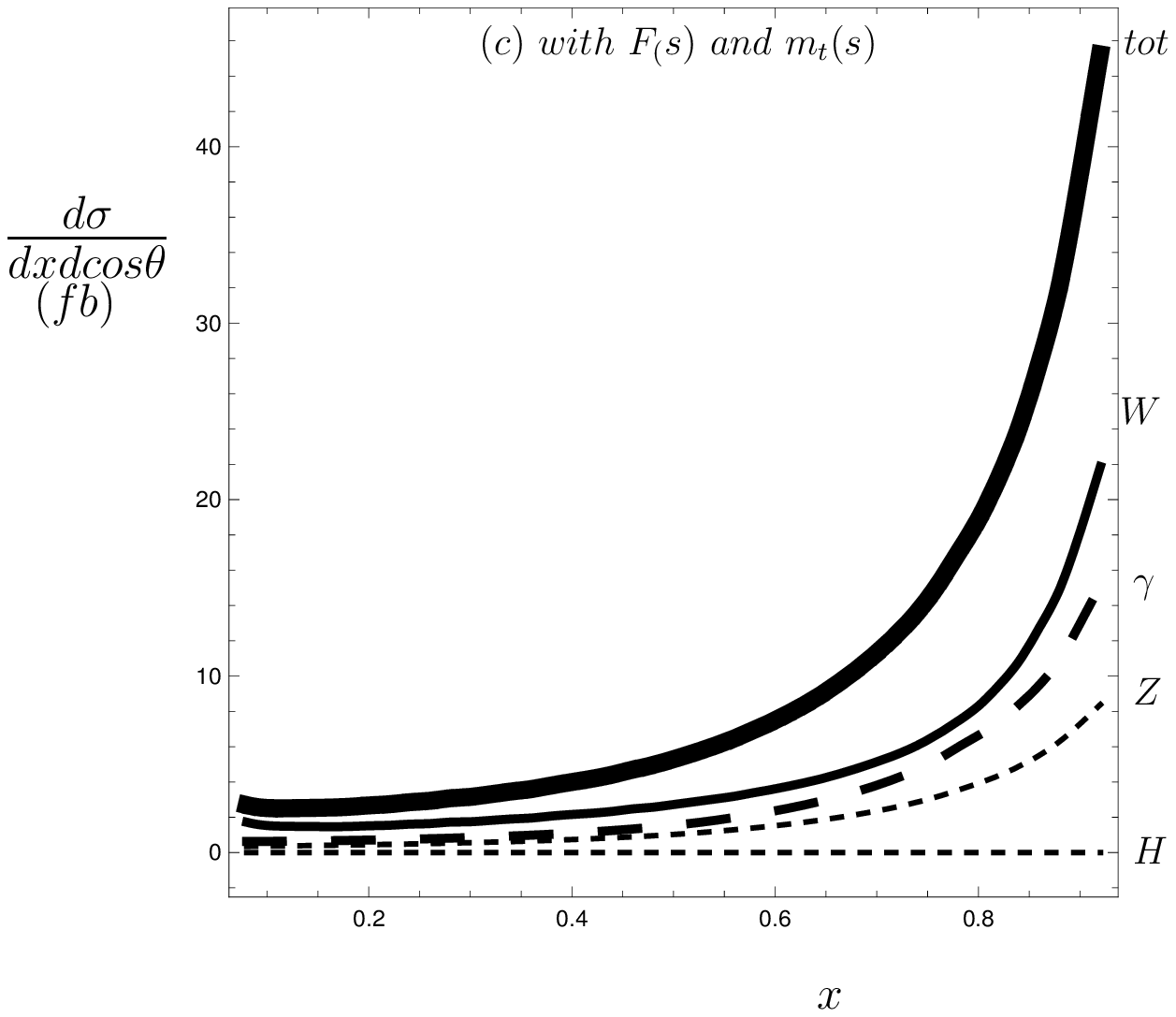, height=10.cm}
\]\\
\vspace{-1cm}
\caption[1] {Same as in Fig. 1 for unpolarized $t$ with effects of $F_R(s)$
(a) and
of $m_t(s)$ (b) in the upper panel
and of both (c) in the lower panel.}
\end{figure}

\clearpage

\begin{figure}[p]
\[\hspace*{-1.8cm}
\epsfig{file=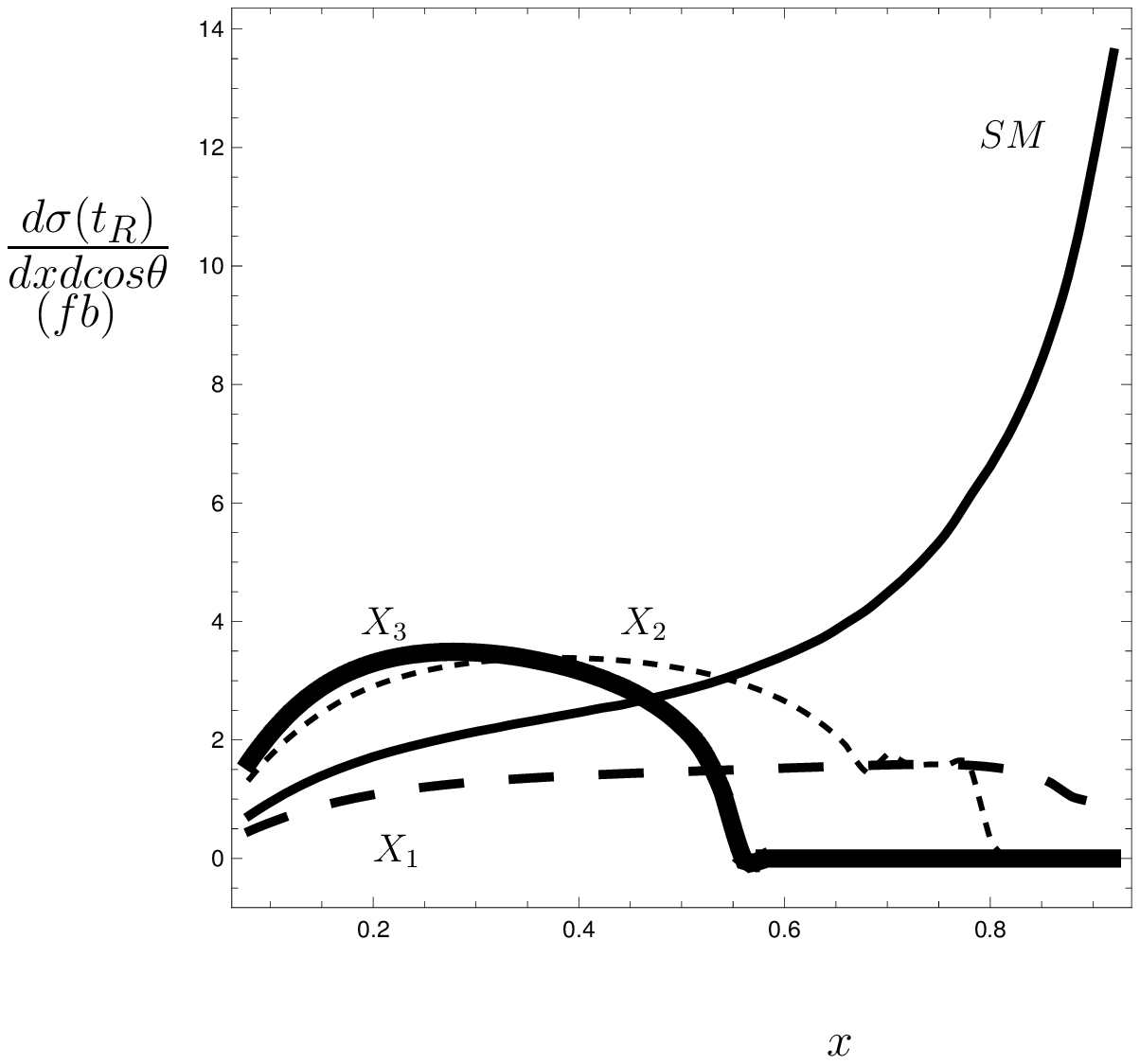, height=7.cm}
\hspace*{1cm}\epsfig{file=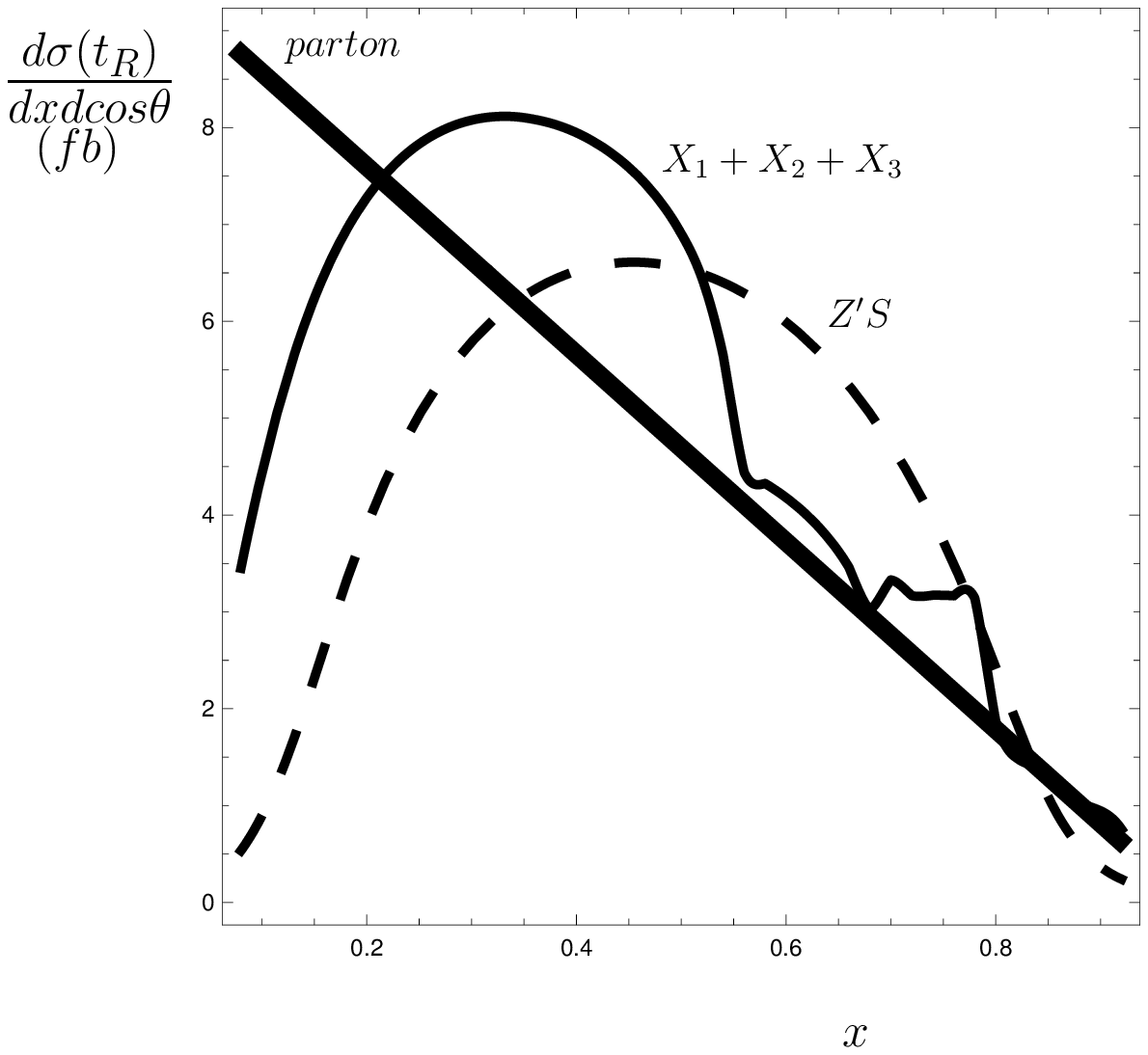, height=7.cm}
\]\\
\[\epsfig{file=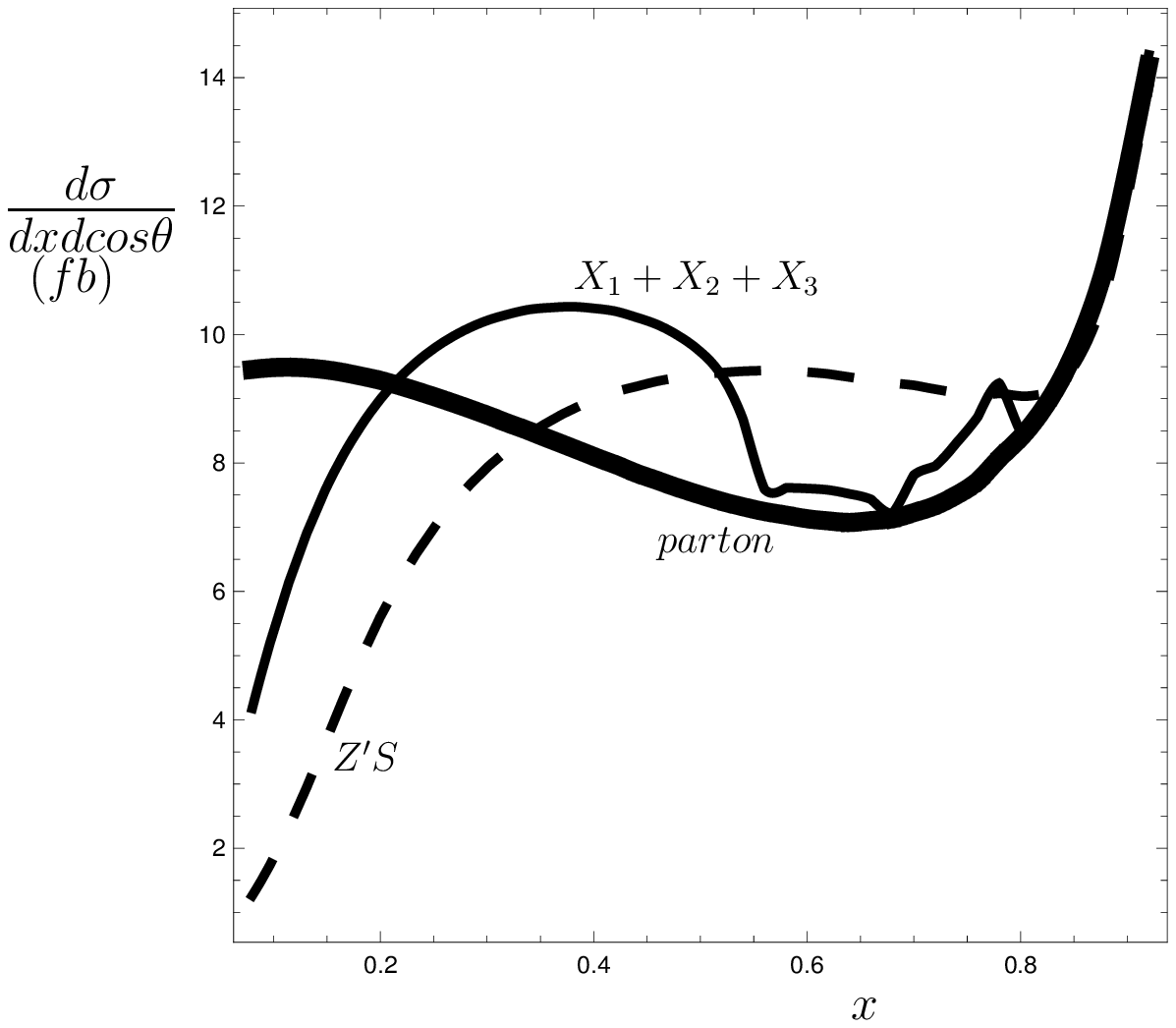, height=7.cm}
\hspace*{1cm}\epsfig{file=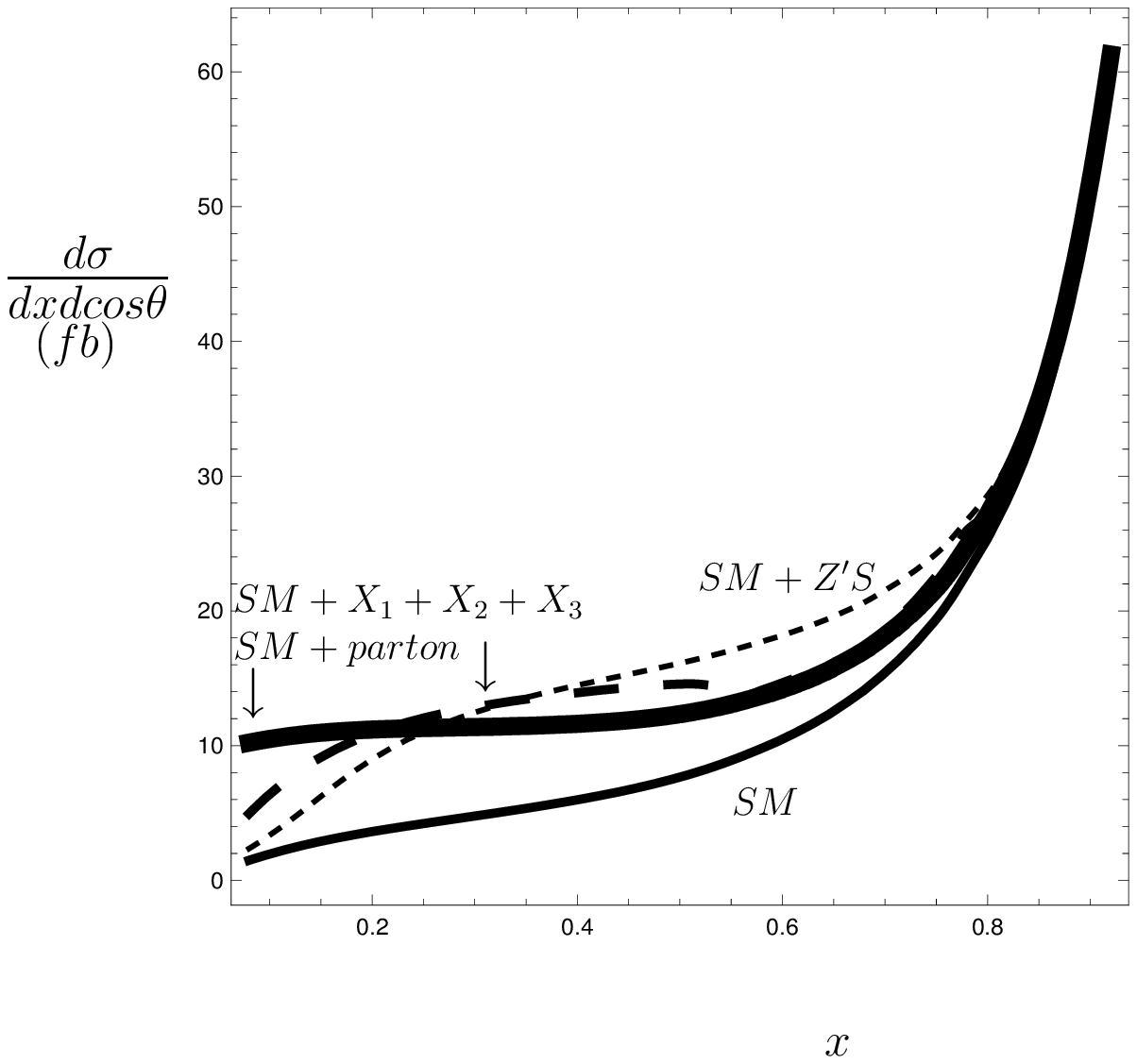, height=7.cm}
\]\\
\vspace{-1cm}
\caption[1] {Results of various additional contributions.
Upper panel for $t_R$: (a) and (b) show individual contributions and their sum
compared to the SM one. Lower panel for unpolarized t: (c) individual contributions, (d) total results when added to SM.}
\end{figure}

\clearpage

\begin{figure}[p]
\[\hspace*{-1.8cm}
\hspace*{1cm}\epsfig{file=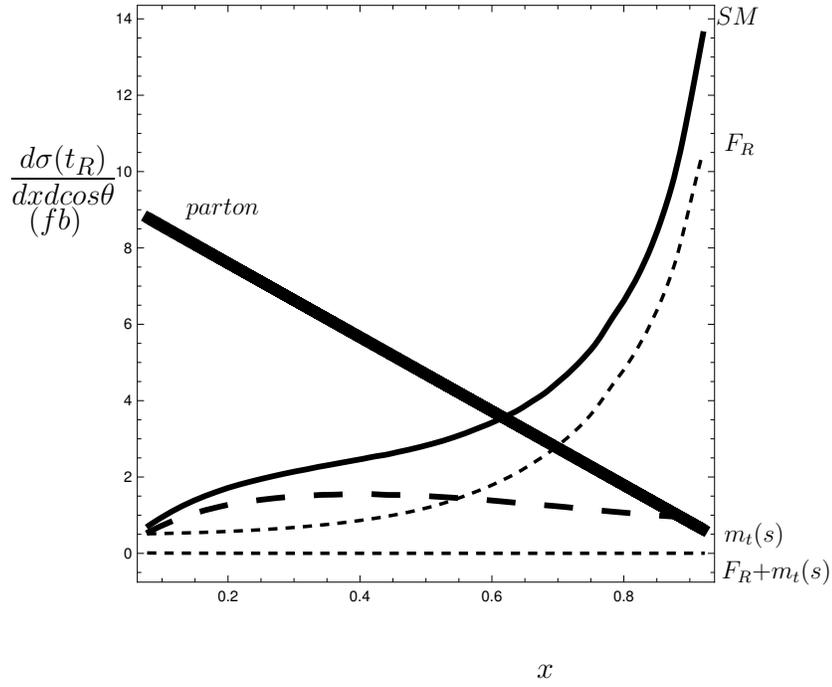, height=9.cm}
\]\\
\[
\epsfig{file=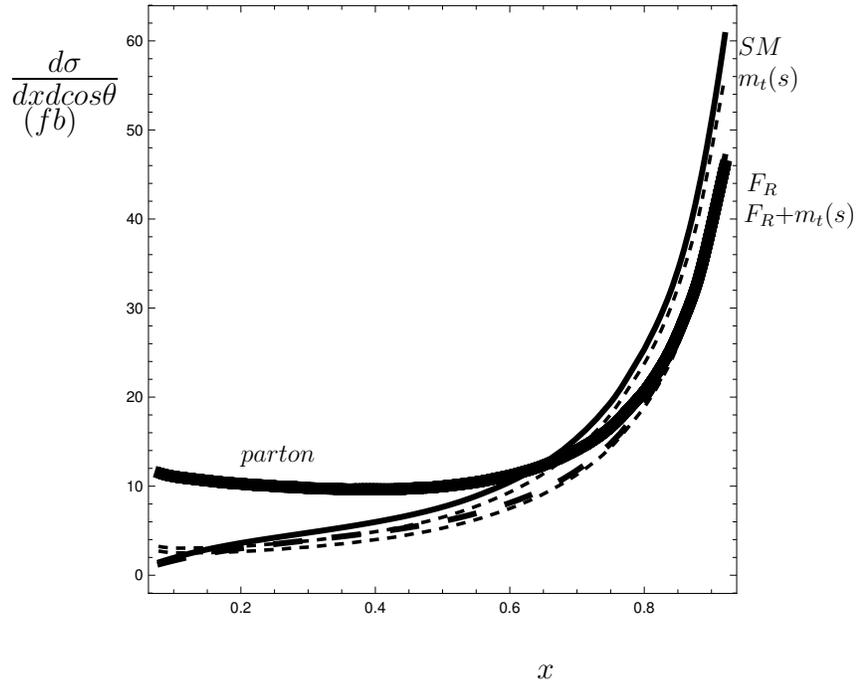, height=9.cm}
\]\\
\vspace{-1cm}
\caption[1] {Resulting effects of $F_R$, of $m_t(s)$, of both and of a parton-like additional contribution; (a) for $t_R$ in the upper panel, (b) for unpolarized t in the lower panel.}
\end{figure}

\clearpage

\end{document}